**Title.** HeartBEiT: Vision Transformer for Electrocardiogram Data Improves Diagnostic Performance at Low Sample Sizes

**Running Title.** HeartBEiT Transformer

**Author Block.** Akhil Vaid MD[1-4], Joy Jiang BS[1-2], Ashwin Sawant MD[5], Stamatios Lerakis MD PhD[6,7], Edgar Argulian MD[6,7], Yuri Ahuja MD[8], Joshua Lampert MD[6,7], Alexander Charney MD[3,9,10], Hayit Greenspan PhD[11], Benjamin Glicksberg PhD[3,4], Jagat Narula MD PhD[6,7], Girish Nadkarni MD MPH[1-4,12]

**Author Affiliations.**
[1]The Charles Bronfman Institute for Personalized Medicine, Icahn School of Medicine at Mount Sinai, New York, New York
[2]Mount Sinai Clinical Intelligence Center, Icahn School of Medicine at Mount Sinai, New York, New York
[3]Department of Genetics and Genomic Sciences, Icahn School of Medicine at Mount Sinai, New York, New York
[4]The Hasso Plattner Institute for Digital Health at Mount Sinai, New York, New York.
[5]Department of Medicine, Icahn School of Medicine at Mount Sinai, New York, New York, USA
[6]Mount Sinai Heart, Icahn School of Medicine at Mount Sinai, New York, NY, USA
[7]Department of Cardiology, Icahn School of Medicine at Mount Sinai, New York, NY, USA
[8]Department of Medicine, NYU Langone Health, New York, NY, USA.
[9]The Pamela Sklar Division of Psychiatric Genomics, Icahn School of Medicine at Mount Sinai, New York, New York.
[10]Department of Psychiatry, Icahn School of Medicine at Mount Sinai, New York, New York.
[11]Department of Biomedical Engineering, Tel Aviv University, Tel Aviv 6997801, Israel.
[12]Division of Nephrology, Department of Medicine, Icahn School of Medicine at Mount Sinai, New York, New York

**Word count.** 3832

**Abstract.** 145 words.
The electrocardiogram (ECG) is a ubiquitous diagnostic modality. Convolutional neural networks (CNNs) applied towards ECG analysis require large sample sizes, and transfer learning approaches result in suboptimal performance when pre-training is done on natural images. We leveraged masked image modeling to create the first vision-based transformer model, HeartBEiT, for electrocardiogram waveform analysis. We pre-trained this model on 8.5 million ECGs and then compared performance vs. standard CNN architectures for diagnosis of hypertrophic cardiomyopathy, low left ventricular ejection fraction and ST elevation myocardial infarction using differing training sample sizes and independent validation datasets. We show that HeartBEiT has significantly higher performance at lower sample sizes compared to other models. Finally, we also show that HeartBEiT improves explainability of diagnosis by highlighting biologically relevant regions of the EKG vs. standard CNNs. Thus, we present the first vision-based waveform transformer that can be used to develop specialized models for ECG analysis especially at low sample sizes.

**Disclosures.** Dr. Nadkarni reports consultancy agreements with AstraZeneca, BioVie, GLG Consulting, Pensieve Health, Reata, Renalytix, Siemens Healthineers and Variant Bio; research funding from Goldfinch Bio and Renalytix; honoraria from AstraZeneca, BioVie, Lexicon, Daiichi Sankyo, Meanrini Health and Reata; patents or royalties with Renalytix; owns equity and stock options in Pensieve Health and Renalytix as a scientific cofounder ; owns equity in Verici Dx;  has received financial compensation as a scientific board member and advisor to Renalytix; serves on the advisory board of Neurona Health; and serves in an advisory or leadership role for Pensieve Health and Renalytix. All other authors have reported that they have no relationships relevant to the contents of this paper to disclose.

**Contact of Corresponding Author.**
Akhil Vaid
Icahn School of Medicine at Mount Sinai, One Gustave L. Levy Place, Box 1243, New York, New York 10029-6500
akhil.vaid@mssm.edu

**Abbreviations.**

| | |
|---|---|
| ECG | Electrocardiogram |
| CNN | Convolutional Neural Network |
| NLP | Natural Language Processing |
| BEiT | Bidirectional Encoder representation from Image Transformers |
| MSHS | Mount Sinai Hospital System |
| LVEF | Left Ventricular Ejection Fraction |
| EHR | Electronic Health Record |
| HCM | Hypertrophic Cardiomyopathy |
| STEMI | ST Elevation Myocardial Infarction |
| MLM | *Masked Language Modeling* |
| MIM | *Masked Image Modeling* |
| AUROC | Area Under the Receiver Operating Curve |
| AURPC | Area Under the Recall Precision Curve |
| GRAD CAM | Gradient-weighted Class Activation Mapping |
| COVID-19 | COronaVIrus Disease 2019 |

# Introduction

The electrocardiogram (ECG) is a body surface level recording of electrical activity within the heart. Owing to its low cost, non-invasiveness, and wide applicability to cardiac disease, the ECG is a ubiquitous investigation and over 100 million ECGs are performed each year within the United States alone[1] in various healthcare settings. However, the ECG is limited in scope since physicians cannot consistently identify patterns representative of disease – especially for conditions which do not have established diagnostic criteria, or in cases when such patterns may be too subtle or chaotic for human interpretation.

Deep learning has been applied to ECG data for several diagnostic and prognostic use cases[2-9]. The vast majority of this work has been built upon Convolutional Neural Networks (CNNs)[10]. Like other neural networks, CNNs are high variance constructs[11], and require large amounts of data to prevent overfitting[12]. CNNs must also be purpose built to accommodate the dimensionality of incoming data, and they have been used for interpreting ECGs both as 1D waveforms and 2D images[13].

In this context, interpreting ECGs as 2D images presents an advantage due to widely available pre-trained models which often serve as starting points for modeling tasks on smaller datasets[14]. This technique is described as *transfer learning* wherein a model that is trained on a larger, possibly unrelated dataset is fine-tuned on a smaller dataset that is relevant to a problem[15]. Transfer learning is especially useful in healthcare since datasets are limited in size due to limited patient cohorts, rarity of outcomes of interest, and costs associated with generating useful labels. As a result, vision models first trained in a supervised manner on natural images[16] often form the basis of models used in healthcare settings. Unfortunately, transfer learning with such natural images is not a universal solution, and it is known to produce suboptimal results when there exist substantial differences in the pre-training and fine-tuning datasets[17].

Transformer based neural networks utilize the *attention* mechanism to establish and define relationships between discrete units of input data known as tokens[18]. A significant benefit that transformers allow for is unsupervised learning from large corpuses of unlabeled data to learn relationships between tokens, and then utilize this information for other downstream tasks[18]. Due to the ease with which unstructured text can be broken down into tokens, transformers have been tremendously successful at Natural Language Processing (NLP) tasks[19-23]. Recent work has extended the functionality of such models into vision-based tasks, leading to the advent of the vision transformer[18, 24, 25].

The first vision transformers were pre-trained on immense labeled datasets and then fine-tuned on smaller datasets to indicate better performance over CNNs at natural image classification[26]. More recently, the *Bidirectional Encoder representation from Image Transformers* (BEiT) approach has allowed large unlabeled datasets to be leveraged for pre-training transformer neural networks[27]. This approach consists of converting parts of an input image, or *patches* into discrete tokens according to the output of a generative model. Such tokens may be considered analogous to the words within a sentence and be used to pre-train a transformer in much the same way as a language model. Since transformers consider global dependencies[28] within inputs, such pre-training may be especially advantageous for ECGs. Certain pathological patterns such as the S1Q3T3 occur in different parts of a recording[29], and a model which considers only contiguous regions may miss them entirely.

We create a novel vision transformer model pre-trained on a large corpus of several million ECGs belonging to a diverse population. We utilize this model to create specialized models for use cases where little data may be available. We then compare performance and saliency maps to baseline models subject to similar constraints.

**Methods**

Data sources

We utilized all available ECG data from five hospitals within the Mount Sinai Health System (MSHS) to pre-train our model. These hospitals (Mount Sinai Hospital, Morningside, West, Beth Israel, and Brooklyn) serve a large patient population that is reflective of the demographic diversity of New York City. ECG data were retrieved from the GE MUSE system for the years 1980-2021 totaling an approximate 8.5 million discrete ECG recordings for 2.1 million patients. ECG data were obtained as structured XML files containing both raw waveforms as well as metadata associated with patient identifiers, time, place, and indication.

For outcome specific fine-tuning of the model, we collected ground-truth labels for the value of the left ventricular ejection fraction (LVEF) from available echocardiogram reports. The modeling task was classification of patients for an LVEF ≤40%, which defines heart failure with reduced ejection fraction.[30] We also collected labels indicative of a diagnosis of Hypertrophic Cardiomyopathy - a genetic disorder wherein the chambers of the heart undergo a pathological increase in thickness resulting in loss of cardiac function and predisposition to fatal arrhythmias.

Finally, we utilized the publicly available PTB-XL dataset for additional external validation. This dataset contains 21,799 ECGs from 18,869 patients from October 1989 - June 1996. These data have been annotated by two cardiologists and contain ground-truth diagnostic labels such as whether an ECG is indicative of a normal recording or changes suggestive of acute ischemia. ECG recordings from this database were used to fine-tune models for detection of ST-Elevation Myocardial Infarction (STEMI). STEMIs are caused by acute loss of blood supply to heart tissue, and can result in a plethora of complications ranging from loss of contractile function to death.

Preprocessing

ECGs utilized within this study each contain waveform data recorded from one of twelve leads, with each lead representing a different perspective on the heart's electrical activity. Both datasets contain ECGs with either 5 or 10 seconds of waveform data per lead sampled at a rate of 500Hz, for a total of 2500 or 5000 samples. The MSHS dataset does not contain data regarding leads III, aVF, aVL, or aVR. However, these leads are derived since they can be re-created from linear transformations of the vectors representing the other leads. In order to maintain uniformity across samples and datasets, all ECGs were truncated to 2500 samples.

We corrected for noise within ECG recordings through application of a *Butterworth bandpass filter* (0.5Hz – 40Hz) followed by the application of a *median filter* on raw waveform data. Processed waveform data so derived was plotted to images with each image containing a total of eight leads. (I, II, and V1 – V6)

Tokens and tokenization

Tokens may be defined as discrete pre-defined sequences which are grouped and analyzed together on a semantic basis. In the context of language modeling, tokens may simply be the words comprising a body of text. The process of separating out data into such discrete sequences and assigning unique numeric identifiers to them is referred to as *Tokenization*[31].

Masked Image Modeling

Transformer based models operate by using self-attention to establish relationships between tokens. A method commonly used for language models is called *Masked Language Modeling* (MLM)[32], wherein a set percentage of the number of tokens input to the model are masked or hidden, and models are pre-trained by having them predict these masked tokens. Collection and labeling of data may be an expensive process, and such costs are amplified for medical datasets. A significant advantage of MLM is that it allows for the usage of large quantities of unlabeled data to pre-train models.

The BEiT approach extends MLM into *Masked Image Modeling* (MIM) wherein 2D input images are separated into patches containing raw pixels which are then used as the tokenized representations of the input image. This tokenization is accomplished using a separately trained image tokenizer with a preset vocabulary that returns a token for each patch. We used the same publicly available image tokenizer (Dall-E) with a vocabulary of 8192 for conversion of ECG images as the original BEiT implementation.

Model selection

We instantiated a 12-layer transformer model with a hidden layer size of 768, and 12 attention heads for a total of approximately 86M parameters. This model, and its downstream derivatives are referred to as "**HeartBEiT**" within the text of this work.

We compared the downstream problem-specific performance of this model to an equivalently sized ImageNet based vision transformer (ViT-B/16: 86M parameters), as well as CNN based approaches common to deep learning as applied to ECGs. These include the largest available pre-trained ResNet model (ResNet-152: 60M parameters) within the *torchvision* library, and a computationally more inexpensive architecture (EfficientNet-B4: 19M parameters) known to demonstrate better performance at image classification despite having fewer parameters. All baselines were pre-trained in a supervised manner on the ImageNet1K dataset containing 1.2M labeled training images.

Pre-training

Input images were resized to 224x224 pixels, but otherwise subject to no other pre-processing. As opposed to natural images, ECG waveforms require maintenance of morphology and order. Random cropping or flipping may lead to loss of information that may only exist within certain segments of an ECG.

Input images were split into square patches of 16 pixels each, for a total of 196 patches per input image **(Figure 1)**. 40% of the input patches were masked for input into the neural network. We used the *AdamW* optimizer with a learning rate of 5e-4. Other training related metrics are as described in Supplementary Methods. The HeartBEiT model was pre-trained on a node consisting of 4 NVIDIA A100-40G GPUs. At approximately 6 hours per epoch, pre-training the model for 300 epochs took around 2.5 months. Model parameters saved at the 300th epoch were used for downstream fine-tuning in all cases **(Supplementary Figure 1).**

Fine-tuning and statistical analysis

Pre-trained models were subjected to a fine-tuning task to demonstrate and compare performance at ECG based classification. We used data from 4 hospitals for detection of LVEF of <40%, and diagnosis of HCM. In either case, the performance of the fine-tuned model was externally validated on data from Morningside hospital. Data from the PTB-XL database were used to fine-tune the pre-trained HeartBEiT model, as well as the two CNNs for detection of STEMI.

Data were separated into a training dataset, an internal testing dataset, and where applicable, an external validation dataset. We modeled conditions of extreme data paucity by reducing training data to either 1%, 10%, 25%, 50% or 100%, and then testing resulting models against common testing data. In all cases, *Group Shuffle Splitting* with a constant random seed was employed to ensure no patients were present in both training and testing data, and that the same patients were part of either dataset across runs.

We set the classification head of each model to a size of two neurons and utilized *CrossEntropy* loss. The *Adam* optimizer on a *OneCycle* learning rate schedule between 3e-4 and 1e-3 over 30 epochs was utilized for fine-tuning and reported performance metrics correspond to the best performance achieved across these epochs. Threshold independent Area Under the Receiver Operating Characteristic curve (AUROC) and Area Under the Precision Recall Curve (AUPRC) metrics were used to calculate and compare model performance. 95% confidence intervals for areas under the curve were generated through 500 iterations of the bootstrap.

Wasserstein Distance

The Wasserstein distance[33] is a metric of the cost required to transform one distribution into another. Given two discrete images, the magnitude of the Wasserstein distance between them is directly proportional to how dissimilar they are. Higher Wasserstein distances between pre-training and fine-tuning data may lead to sub-optimal results with transfer learning.

We randomly sampled 1000 images each from both the ImageNet and ECG datasets. All samples from within each cohort were resized to 224x224 pixels and paired against all other samples from the same cohort, as well as the other cohort for a total of 3 such combinations: ECG vs ECG, ECG vs ImageNet, ImageNet vs ImageNet. Each such operation yielded a total of $10^6$ pairs. The Wasserstein distance was calculated for each resulting pair of images and averaged across the combination of cohorts.

Explainability

Model explainability was generated using the *FullGrad* implementation available within the *Gradient-weighted Class Activation Mapping* (GradCAM*)* library[34]. Generated tensor attribution scores were plotted as an overlay upon the original input image to demonstrate which part of an input contributed most to a prediction.

Software

All analyses were performed using the *pandas, numpy, Python Image Library (PIL), SciPy, scikit-learn, torchvision, timm,* and *PyTorch* libraries. Plotting was performed using the *matplotlib* and *seaborn* libraries. All code was written for and within the 3.8.x version of the Python programming language.

# Results

Performance at classification of LVEF

We included 511,491 total ECGs from MSHS in the training or fine-tuning set, 20,448 samples from MSHS in testing, and 1,480 from Morningside in external validation. Low LVEF prevalence was 18% in the training set **(Table 1).**

HeartBEiT outperformed other CNN models at low LVEF classification at all fractions of training data **(Figure 2; Table 2).** At 1% of training data (5,114 samples), performance (AUROC: 0.86, 95% CI: 0.86-0.86) was 28.4% better than the ViT-B/16 model (AUROC: 0.67, 95% CI 0.67-0.67), 5.2% better than EfficientNet-B4 (AUROC: 0.82, 95% CI: 0.82-0.82), and 2.4% better than ResNet-152 (AUROC: 0.84, 95% CI: 0.84-0.84) in internal testing (Supplementary Figure 2). These trends were maintained across external validation with HeartBEiT (AUROC: 0.87, 95% CI: 0.87-0.87) outperforming the CNNs by 4-18% (**Supplementary Figure 4**).

Using AUPRC as a metric, at 1% of training data and against a prevalence of 18.5% in the internal testing cohort, the HeartBEiT model (AUPRC: 0.59, 95% CI: 0.59-0.59) outperformed Vit-B/16 (AUPRC: 0.31, 95% CI 0.31-0.31) by 90.3%, EfficientNet-B4 (AUPRC: 0.48, 95% CI: 0.48-0.48) by 22.9% and the ResNet-152 (AUPRC: 0.52, 95% CI: 0.52-0.52) by 13.5% **(Figure 2; Table 3; Supplementary Table 3).** In the external validation cohort, HeartBEiT had the highest AUPRC of 0.73 (95% CI: 0.73-0.73) **(Supplementary Table 5)**.

With 100% of the training data (511,491 samples), performance across all models became more closely matched. In internal testing, there was no performance differential among HeartBEiT, EfficientNet, and ResNet, and a differential of 1.1-4.5% was observed in external validation for AUROC. However, for AUPRC, HeartBEiT still had improved performance of 0-17.7% in internal and external datasets.

GRAD-CAM analysis demonstrated areas around the QRS complexes of each lead were highlighted at 1% of training data by HeartBEiT **(Supplementary Figure 6a).** When 100% of training data were implemented, foci became more pronounced around the QRS complexes of lead I **(Supplementary Figure 6b).**

Performance at diagnosis of HCM

We fine-tuned the HeartBEiT transformer using 78,831 ECGs from four hospitals of the MSHS. Testing was conducted on 20,448 ECGs from these hospitals, and 3,859 ECGs from a holdout set of patients from Morningside were used for external validation **(Table 1)**. The prevalence of HCM in the training set was 38%.

HeartBEiT outperformed the other models at diagnosis of HCM at all fractions of training data **(Figure 3; Table 2)**. At 1% of training data, performance of the HeartBEiT model at AUROC of 0.77 (95% CI: 0.77-0.77) exceeded that of Vit-B/16 by 26.2% and those of both EfficientNet and ResNet by 6.9% in internal testing **(Supplementary Figure 2)**. Similar results were seen for external validation with the HeartBEiT model which had an AUROC of 0.74 (95% CI: 0.74-0.74), outperforming Vit-B/16 (0.61, 95% CI 0.61-0.61) by 21.3%, EfficientNet-B4 (0.69, 95% CI: 0.68-0.70) by 7.2%, and ResNet-152 (0.68, 95% CI: 0.68-0.69) by 8.8% **(Supplementary Table 4)**.

Differences in performance were much more profound for AUPRC at 1% of training data in use **(Figure 3; Table 3)**. Using 1% of training data, against an outcome prevalence of 38.8% in the internal testing cohort, the HeartBEiT model (AUPRC: 0.67 95%, CI: 0.67-0.67) exceeded performance of Vit-B/16 (AUPRC: 0.50, 95% CI 0.50-0.50) by 34.0%, EfficientNet-B4 (AUPRC: 0.63, 95% CI: 0.63-0.63) by 6.3% and the ResNet-152 (AUPRC: 0.64, 95% CI: 0.64-0.64) by 4.7% **(Supplementary Figure 3)**. In external validation, HeartBEiT continued to exhibit the best performance with AUPRC of 0.64 (95% CI: 0.64-0.64) **(Supplementary Figure 5)**.

The HeartBEiT performance advantage reduced gradually as the amount of training data increased. Compared to 100% of the training data, the performance differential was up to 2.5% in internal testing and 3.9% external validation for AUROC and up to 4.2% and 7.1% for internal testing and external validation, respectively, for AUPRC.

GRAD-CAM analysis revealed that at 1% of the data, the QRS complexes of lead I, V2, and V5 and the ST segment of V6 were denoted as important regions for predicting HCM by HeartBEiT **(Supplementary Figure 7a)**. In contrast, at 100% of the training data, key areas identified by HeartBEiT became more focused to the beginning of V5 **(Supplementary Figure 7b)**.

Performance at detection of STEMI

The PTB-XL dataset contains 21,799 total ECGs from 18,869 patients: 17,449 ECGs were used fine-tuning and 4,352 to test the model. The prevalence of STEMI was around 5.7% in the training set and 5.4% in the testing set **(Table 1)**.

The AUROC performance advantage of HeartBEiT was seen to be greater at smaller fractions of training data used for training. In internal testing, the AUROC of HeartBEiT was 0.88 (95% CI: 0.88-0.89) with 4.8-10% performance improvement compared to the other models at 1% of training data **(Figure 4; Table 2; Supplementary Table 2)**. This advantage changed to approximately 20.3%, 1.1%, and 2.2% in comparison to ViT-B/16, EfficientNet, and ResNet, respectively, when all available training data (17,449 samples) were used.

This performance advantage became much more profound for AUPRC, with HeartBEiT (AUPRC: 0.56, 95% CI 0.56-0.66) outperforming Vit-B/16 (0.27, 95% CI 0.26-37) by 107.4%, ResNet (0.47, 95% CI 0.46-0.47) by 19.1% and the EfficientNet (0.40, 95% CI 0.40-0.41) by 40.0% at a 1% fraction of training data **(Table 3; Supplementary Table 3).** However, at 100% of training data, performance of HeartBEiT (AUPRC: 0.67, 95% CI 0.66-0.67) became non-significantly lower than that of EfficientNet (AUPRC: 0.68, 95% CI: 0.67-0.68).

For STEMI detection, the ViT-B/16 vision transformer exhibited training instability when using more than 10% of training data while keeping other hyperparameters such as learning rate constant. This instability was seen only for this outcome, and reported performance corresponds to best metrics achieved prior to the training methods erroring out.

ST segments in each lead were underscored as areas of importance according to GRAD-CAM analysis of HeartBEiT at 1% of the training data **(Figure 5).** At 100% of the training data, these areas denoted by HeartBEiT became localized around ST segments of leads V3 and V4 **(Supplementary Figure 8)**.

Wasserstein Distance

The average pairwise Wasserstein distance for the ECG vs ECG set was 2.14. In comparison, this value was 45.48 for the ImageNet vs ImageNet set, and 128.44 for the ECG vs ImageNet set **(Supplementary Figure 9).**

## Discussion

Using 8.5 million ECGs from 2.1 million patients collected over a period of four decades, we leveraged Masked Image Modeling to create the first vision-based transformer (HeartBEiT) model for ECG data that can act as a universal starting point for downstream training on outcomes of interest. We fine-tuned this model against two outcomes using data derived from four hospitals within the Mount Sinai Health System, and externally validated derived models on data from another hospital. We also fine-tuned this model for STEMI detection using data from the publicly available PTB-XL database, followed by testing the derived model against a holdout set of patients. In each case, our model was compared against two CNNs all subject to the same training conditions. Finally, we evaluated an additional aspect of clinical usefulness of these models by creating saliency maps for input samples.

Neural network performance can be heavily influenced by the amount of data available[35], and overfitting can easily result in small data regimes[36]. However, curated labeled data is a scarce resource. This is especially true in the healthcare setting wherein performing testing on patients, detecting pathologies of interest, and gathering data regarding clinical outcomes is laborious and expensive. In addition to the financial costs of acquiring and labelling data, time may be an additional factor that precludes acquisition of larger datasets. During emergent public health concerns, such as the recent COVID-19 pandemic, little data may be available for the development of useful models. In such circumstances, models that can work with a fraction of the data required for other approaches may assist in quicker, more appropriate diagnosis and triage.

Across all outcomes, datasets, and performance metrics, we found that HeartBEiT required 10% of the training data to achieve the performance equivalent of using 100% of the training data. Further, in the very low data regime using only 1% of training data, HeartBEiT performance was equivalent to other models using 10 times as much data. This performance was maintained in external validation not only for the fine-tuned models, but also for the pre-trained model when used with an altogether new dataset from an independent dataset comprised of a geographically separated cohort of patients.

Of special importance is the elevated difference in performance in the AUPRC - a better indicator of performance in datasets with heavy class imbalance wherein considering AUROC in isolation may be less useful. Given relatively low event rates, medical datasets tend to have such class imbalances. For example, in detection of STEMI with an outcome prevalence of 5.6%, in the 1% training data regime, HeartBEiT exceeded the AUPRC of the CNNs by 19.1% and 40% respectively, while doubling the performance of the ImageNet vision transformer. These results also indicate that pre-training on natural images isn't always the most optimal solution for creating healthcare related models – a fact further evidenced by the extent of the disparity in the average Wasserstein distance between natural images and ECGs.

An emergent clinical advantage of using transformers with the explainability framework described in this work is the granularity of the saliency mapping. Even at similar levels of performance, the CNNs shown tend to coalesce areas of importance, thereby obfuscating the strongest determinants of a prediction. In comparison, saliency maps for transformers tend to focus on these determinants. Such granular explainability may help both clinician adoption of deep learning models[37], as well as aid in understanding pathologies for which there are no diagnostic guidelines on an ECG. These factors are demonstrated well for STEMI detection where the pathognomonic pattern is well established, and the ST segment is consistently highlighted even when using 1% of data for fine-tuning **(Figure 5)**. In the case of LVEF determination, there exist no clear diagnostic guidelines that can assist human physicians. In this case, saliency maps tend to focus on QRS complexes which indicate the net vector of depolarization of the majority of the cardiac ventricular musculature and point towards the transformer's ability to focus on the mechanisms underlying the disease condition.

Our work must be considered in light of certain limitations. Transformers tend to be very compute intensive to pre-train. We were therefore limited in the size of the transformer model at 86M parameters, as well as the dimensions of the input data we were able to utilize. However, we believe this work serves as evidence of the viability and advantages of our HeartBEiT model, and future work will deal with scaling up this model to enable better performance prior to live deployment.

In conclusion, pre-trained transformer models enable robust deep learning-based ECG classification even in severely data limited regimes. More specific, better quality, granular saliency maps can aid clinician acceptance of model predictions.

# Figures

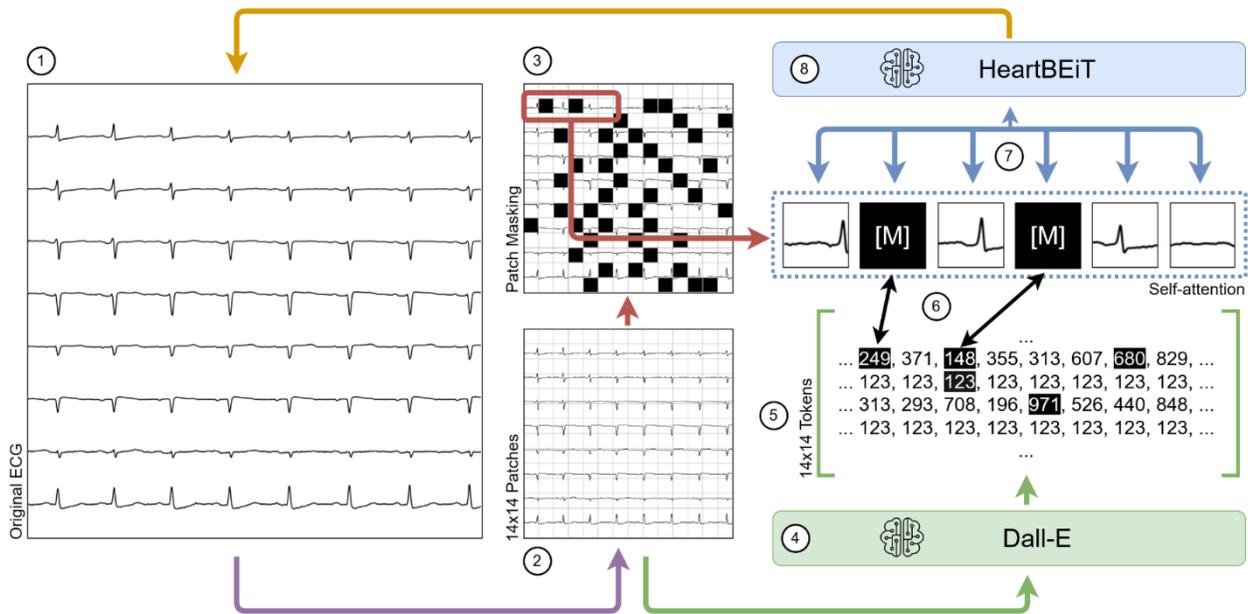

**Figure 1. Pre-training**
Pre-training of the HeartBEiT model. (1) Each original ECG is partitioned into patches (2) of 14x14 pixels. These patches are tokenized, and some of them are masked (3). The Dall-E model (4) acts as the tokenizer and converts the image into discrete tokens (5) which are then made part of the Masked Image Modeling process (6). This allows for pre-training the HeartBEiT model's attention modules (7), and the model is then used for downstream fine-tuning (8).

# Left Ventricular Ejection Fraction <= 40%

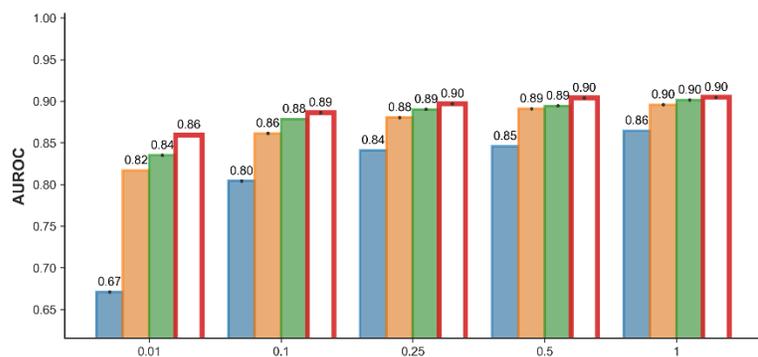
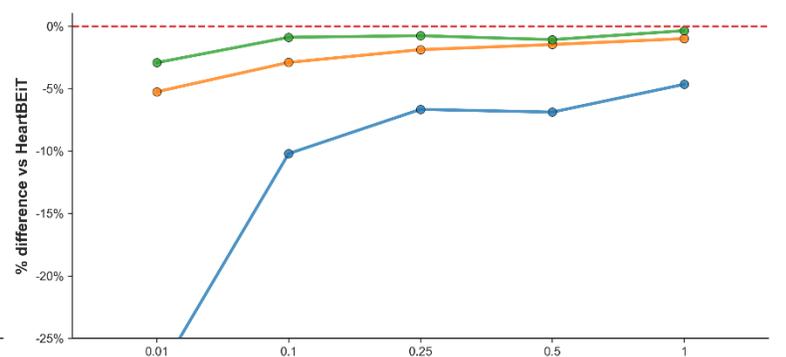
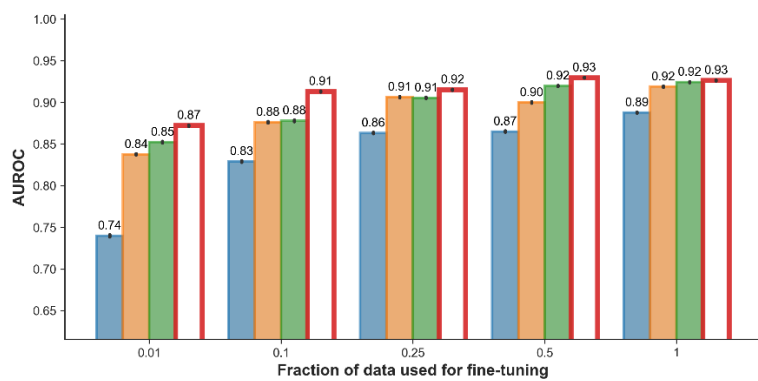
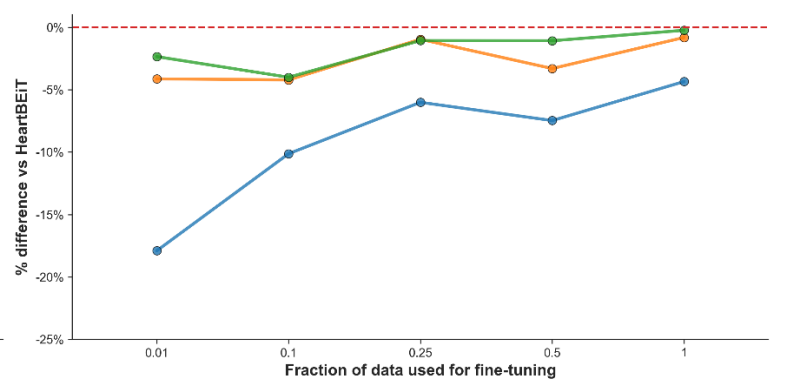

**Figure 2. Left Ventricular Ejection Fraction <= 40% classification on ECGs**
Panel A. Internal testing performance (4 Mount Sinai facilities)
Panel B. Internal testing performance difference
Panel C. External validation performance (Morningside patients)
Panel D. External validation performance difference

# Hypertrophic Cardiomyopathy

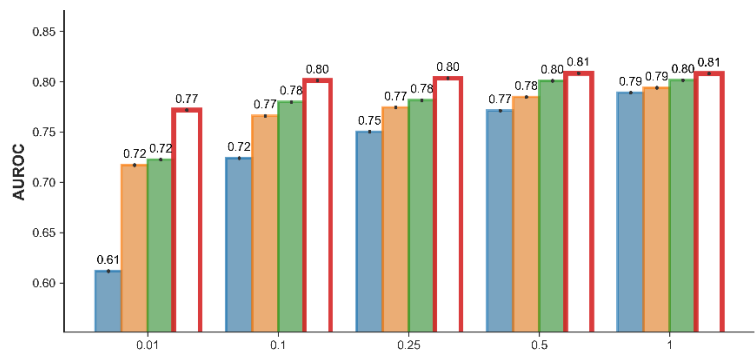
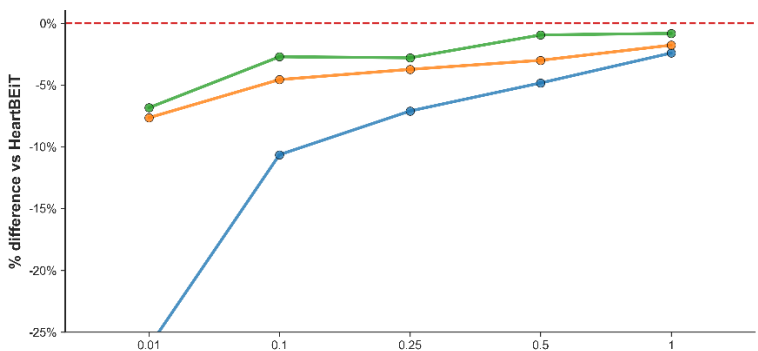
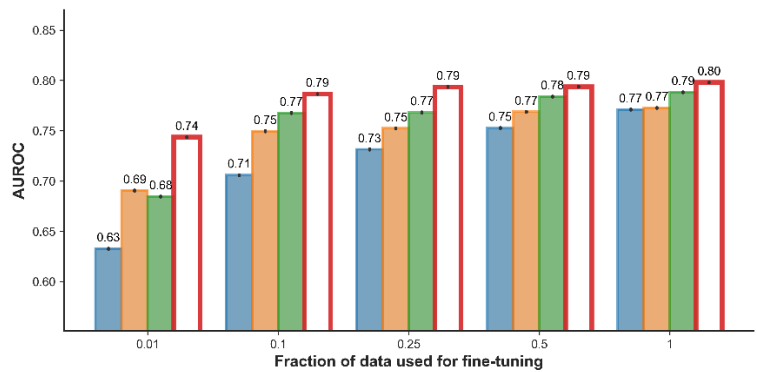
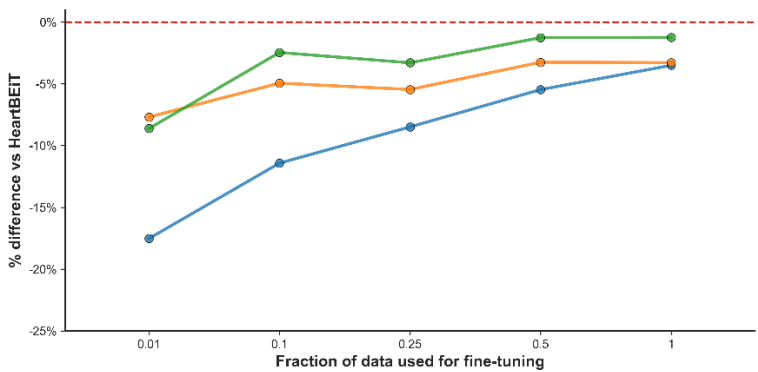

**Figure 3. Hypertrophic cardiomyopathy classification on ECGs**
Panel A. Internal testing performance (4 Mount Sinai facilities)
Panel B. Internal testing performance difference
Panel C. External validation performance (Morningside patients)
Panel D. External validation performance difference

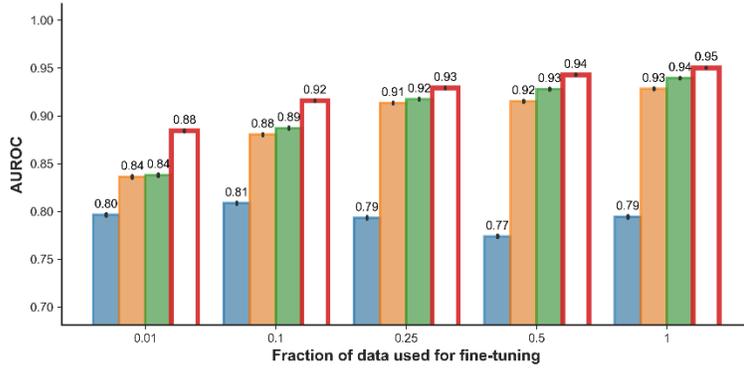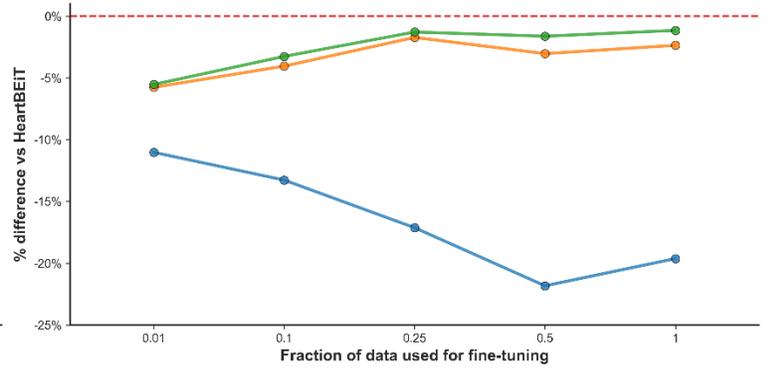

**Figure 4. STEMI detection on ECGs (PTB-XL database)**
A. Internal testing performance
B. Internal testing performance difference

# ST-Elevation Myocardial Infarction

Fraction of training data: 0.01

**A** ViT-B/16

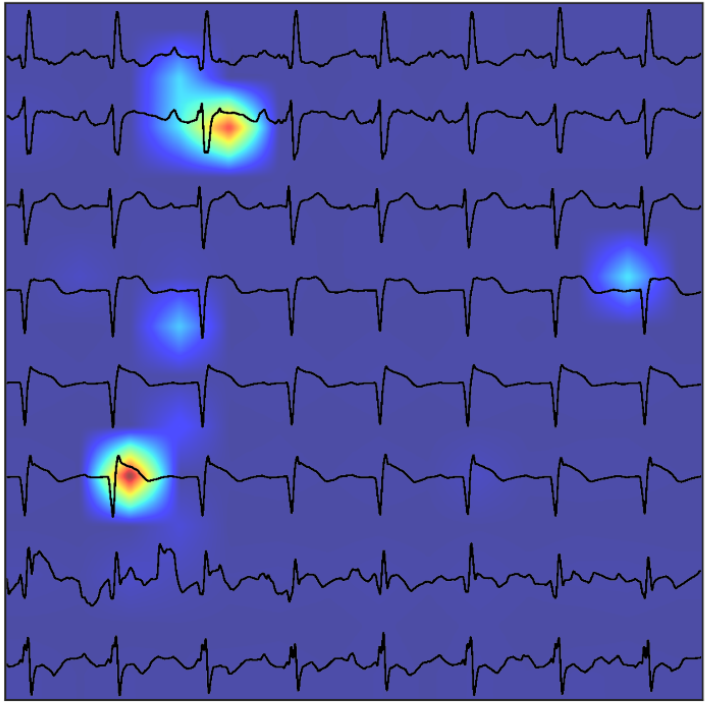

**B** EfficientNet-B4

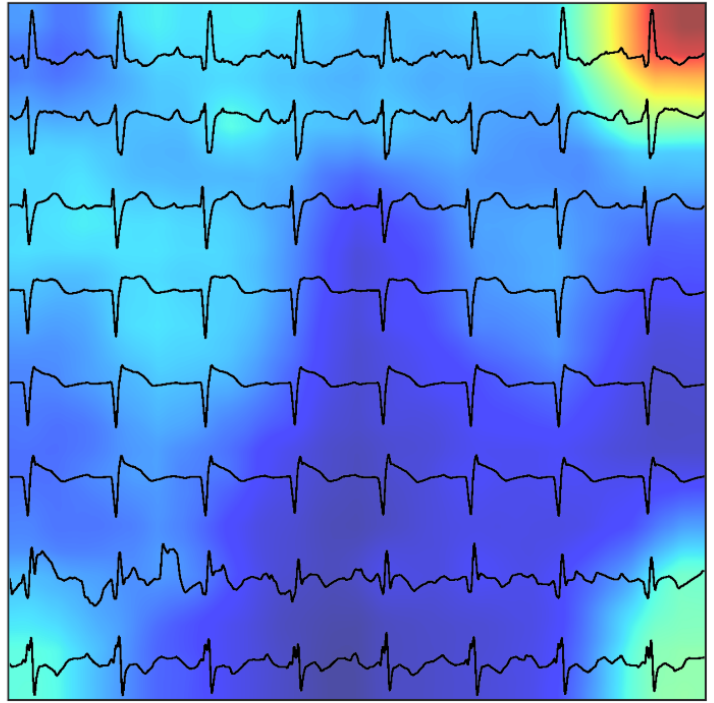

**C** ResNet152

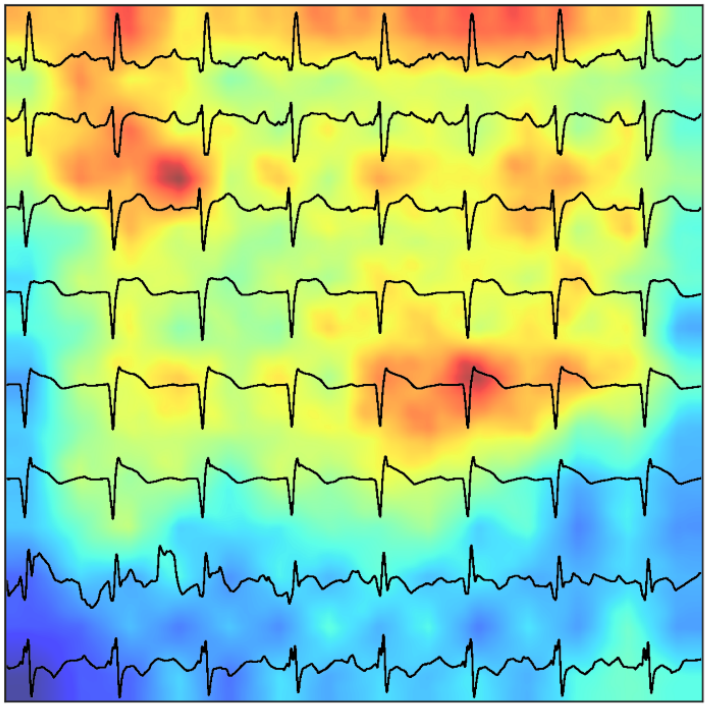

**D** HeartBEiT

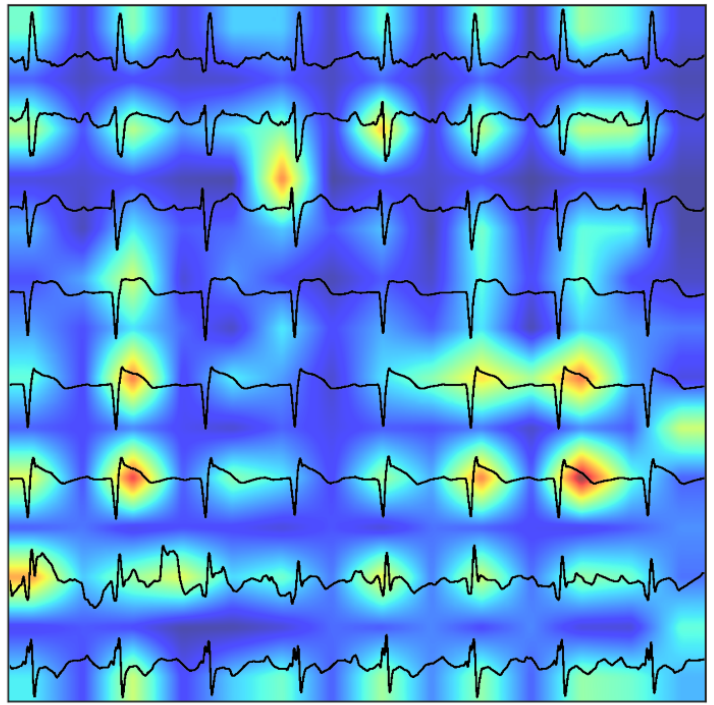

**Figure 5. Saliency mapping for STEMI detection at 1% training data**

Panel A. ViT-B/16
Panel B. EfficientNet-B4
Panel C. ResNet-152
Panel D. HeartBEiT

HeartBEiT localizes to the ST segments. Other models are more diffuse in highlighting features of importance and may be less useful clinically.

**Tables**

|  | Fine-tuning | Testing | External Validation |
|---|---:|---:|---:|
| **Low LVEF** | | | |
| Number of ECGs (n) | 511,491 | 128,687 | 1,480 |
| Outcome Prevalence (%) | 18.4 | 18.6 | 26.6 |
| **Hypertrophic Cardiomyopathy** | | | |
| Number of ECGs (n) | 78,831 | 20,448 | 13,859 |
| Outcome Prevalence (%) | 37.4 | 38.8 | 36.6 |
| **STEMI (PTB-XL database)** | | | |
| Number of ECGs (n) | 17,449 | 4,352 | - |
| Outcome Prevalence (%) | 5.7 | 5.4 | - |

**Table 1.** Dataset Size with Outcome Prevalence

|  | Model | 0.01 | 0.1 | 0.25 | 0.5 | 1 |
|---|---|---|---|---|---|---|
| | **Left ventricular ejection fraction <= 40%** | | | | | |
| Internal testing | ViT-B/16 | 0.67 (0.67 - 0.67) | 0.80 (0.80 - 0.80) | 0.84 (0.84 - 0.84) | 0.85 (0.85 - 0.85) | 0.86 (0.86 - 0.86) |
| | EfficientNet-B4 | 0.82 (0.82 - 0.82) | 0.86 (0.86 - 0.86) | 0.88 (0.88 - 0.88) | 0.89 (0.89 - 0.89) | 0.90 (0.90 - 0.90) |
| | ResNet-152 | 0.84 (0.84 - 0.84) | 0.88 (0.88 - 0.88) | 0.89 (0.89 - 0.89) | 0.89 (0.89 - 0.89) | 0.90 (0.90 - 0.90) |
| | **HeartBEiT** | **0.86 (0.86 - 0.86)** | **0.89 (0.89 - 0.89)** | **0.90 (0.90 - 0.90)** | **0.90 (0.90 - 0.90)** | **0.90 (0.90 - 0.90)** |
| | Fine-tuning samples | 5,114 | 51,149 | 127,872 | 255,745 | 511,491 |
| | Testing samples | 128,687 | 128,687 | 128,687 | 128,687 | 128,687 |
| External validation | ViT-B/16 | 0.74 (0.74 - 0.74) | 0.83 (0.83 - 0.83) | 0.86 (0.86 - 0.86) | 0.87 (0.86 - 0.87) | 0.89 (0.89 - 0.89) |
| | EfficientNet-B4 | 0.84 (0.84 - 0.84) | 0.88 (0.88 - 0.88) | 0.91 (0.91 - 0.91) | 0.90 (0.90 - 0.90) | 0.92 (0.92 - 0.92) |
| | ResNet-152 | 0.85 (0.85 - 0.85) | 0.88 (0.88 - 0.88) | 0.91 (0.90 - 0.91) | 0.92 (0.92 - 0.92) | 0.92 (0.92 - 0.92) |
| | **HeartBEiT** | **0.87 (0.87 - 0.87)** | **0.91 (0.91 - 0.91)** | **0.92 (0.91 - 0.92)** | **0.93 (0.93 - 0.93)** | **0.93 (0.93 - 0.93)** |
| | Testing samples | 1,480 | 1,480 | 1,480 | 1,480 | 1,480 |
| | **Hypertrophic Cardiomyopathy** | | | | | |
| Internal testing | ViT-B/16 | 0.61 (0.61 - 0.61) | 0.72 (0.72 - 0.72) | 0.75 (0.75 - 0.75) | 0.77 (0.77 - 0.77) | 0.79 (0.79 - 0.79) |
| | EfficientNet-B4 | 0.72 (0.72 - 0.72) | 0.77 (0.77 - 0.77) | 0.77 (0.77 - 0.77) | 0.78 (0.78 - 0.79) | 0.79 (0.79 - 0.79) |
| | ResNet-152 | 0.72 (0.72 - 0.72) | 0.78 (0.78 - 0.78) | 0.78 (0.78 - 0.78) | 0.80 (0.80 - 0.80) | 0.80 (0.80 - 0.80) |
| | **HeartBEiT** | **0.77 (0.77 - 0.77)** | **0.80 (0.80 - 0.80)** | **0.80 (0.80 - 0.80)** | **0.81 (0.81 - 0.81)** | **0.81 (0.81 - 0.81)** |
| | Fine-tuning samples | 788 | 78,83 | 19,707 | 39,415 | 78,831 |
| | Testing samples | 20,448 | 20,448 | 20,448 | 20,448 | 20,448 |
| External validation | ViT-B/16 | 0.63 (0.63 - 0.63) | 0.71 (0.71 - 0.71) | 0.73 (0.73 - 0.73) | 0.75 (0.75 - 0.75) | 0.77 (0.77 - 0.77) |
| | EfficientNet-B4 | 0.69 (0.69 - 0.69) | 0.75 (0.75 - 0.75) | 0.75 (0.75 - 0.75) | 0.77 (0.77 - 0.77) | 0.77 (0.77 - 0.77) |
| | ResNet-152 | 0.68 (0.68 - 0.68) | 0.77 (0.77 - 0.77) | 0.77 (0.77 - 0.77) | 0.78 (0.78 - 0.78) | 0.79 (0.79 - 0.79) |
| | **HeartBEiT** | **0.74 (0.74 - 0.74)** | **0.79 (0.79 - 0.79)** | **0.79 (0.79 - 0.79)** | **0.79 (0.79 - 0.79)** | **0.80 (0.80 - 0.80)** |
| | Testing samples | 13,859 | 13,859 | 13,859 | 13,859 | 13,859 |
| | **ST-Elevation Myocardial Infarction** | | | | | |
| Internal testing | ViT-B/16 | 0.80 (0.80 - 0.80) | 0.81 (0.81 - 0.81) | 0.79 (0.79 - 0.79) | 0.77 (0.77 - 0.78) | 0.79 (0.79 - 0.80) |
| | EfficientNet-B4 | 0.84 (0.83 - 0.84) | 0.88 (0.88 - 0.88) | 0.91 (0.91 - 0.91) | 0.92 (0.91 - 0.92) | 0.93 (0.93 - 0.93) |
| | ResNet-152 | 0.84 (0.84 - 0.84) | 0.89 (0.89 - 0.89) | 0.92 (0.92 - 0.92) | 0.93 (0.93 - 0.93) | 0.94 (0.94 - 0.94) |

| | HeartBEiT | 0.88<br>(0.88 - 0.89) | 0.92<br>(0.92 - 0.92) | 0.93<br>(0.93 - 0.93) | 0.94<br>(0.94 - 0.94) | 0.95<br>(0.95 - 0.95) |
|---|---|---|---|---|---|---|
| | Fine-tuning samples | 174 | 1,744 | 4,362 | 8,724 | 17,449 |
| | Testing samples | 4,352 | 4,352 | 4,352 | 4,352 | 4,352 |

**Table 2.** Area Under the Receiver Operating Characteristic Curve (AUROC) metrics for various models compared to HeartBEiT at different fractions of fine-tuning and testing data

|  | Model | 0.01 | 0.1 | 0.25 | 0.5 | 1 |
|---|---|---|---|---|---|---|
| | **Left ventricular ejection fraction <= 40%** | | | | | |
| Internal testing | ViT-B/16 | 0.31 (0.31 - 0.31) | 0.46 (0.46 - 0.46) | 0.55 (0.55 - 0.55) | 0.57 (0.57 - 0.57) | 0.62 (0.62 - 0.62) |
| | EfficientNet-B4 | 0.48 (0.48 - 0.48) | 0.62 (0.62 - 0.62) | 0.67 (0.67 - 0.67) | 0.70 (0.70 - 0.70) | 0.71 (0.71 - 0.71) |
| | ResNet-152 | 0.52 (0.52 - 0.52) | 0.65 (0.65 - 0.65) | 0.69 (0.69 - 0.69) | 0.70 (0.70 - 0.70) | 0.72 (0.72 - 0.72) |
| | **HeartBEiT** | **0.59 (0.59 - 0.59)** | **0.68 (0.68 - 0.68)** | **0.71 (0.71 - 0.71)** | **0.73 (0.73 - 0.73)** | **0.73 (0.73 - 0.73)** |
| | Fine-tuning samples | 5,114 | 51,149 | 127,872 | 255,745 | 511,491 |
| | Testing samples | 128,687 | 128,687 | 128,687 | 128,687 | 128,687 |
| External validation | ViT-B/16 | 0.49 (0.48 - 0.49) | 0.66 (0.66 - 0.66) | 0.70 (0.70 - 0.70) | 0.73 (0.72 - 0.73) | 0.74 (0.74 - 0.74) |
| | EfficientNet-B4 | 0.65 (0.65 - 0.65) | 0.76 (0.76 - 0.76) | 0.82 (0.82 - 0.82) | 0.82 (0.82 - 0.82) | 0.84 (0.84 - 0.85) |
| | ResNet-152 | 0.67 (0.67 - 0.67) | 0.77 (0.76 - 0.77) | 0.82 (0.82 - 0.82) | 0.84 (0.84 - 0.84) | 0.85 (0.85 - 0.85) |
| | **HeartBEiT** | **0.73 (0.73 - 0.73)** | **0.83 (0.83 - 0.83)** | **0.83 (0.83 - 0.84)** | **0.86 (0.86 - 0.86)** | **0.85 (0.85 - 0.85)** |
| | Testing samples | 1,480 | 1,480 | 1,480 | 1,480 | 1,480 |
| | **Hypertrophic Cardiomyopathy** | | | | | |
| Internal testing | ViT-B/16 | 0.49 (0.49 - 0.49) | 0.62 (0.62 - 0.62) | 0.64 (0.64 - 0.65) | 0.68 (0.68 - 0.68) | 0.71 (0.71 - 0.71) |
| | EfficientNet-B4 | 0.63 (0.63 - 0.63) | 0.68 (0.68 - 0.68) | 0.68 (0.68 - 0.68) | 0.70 (0.70 - 0.70) | 0.72 (0.72 - 0.72) |
| | ResNet-152 | 0.64 (0.64 - 0.64) | 0.69 (0.69 - 0.69) | 0.70 (0.70 - 0.70) | 0.72 (0.72 - 0.72) | 0.72 (0.72 - 0.72) |
| | **HeartBEiT** | **0.67 (0.67 - 0.67)** | **0.72 (0.72 - 0.72)** | **0.73 (0.73 - 0.73)** | **0.74 (0.74 - 0.74)** | **0.74 (0.73 - 0.74)** |
| | Fine-tuning samples | 788 | 78,83 | 19,707 | 39,415 | 78,831 |
| | Testing samples | 20,448 | 20,448 | 20,448 | 20,448 | 20,448 |
| External validation | ViT-B/16 | 0.50 (0.50 - 0.50) | 0.59 (0.59 - 0.59) | 0.63 (0.63 - 0.63) | 0.68 (0.67 - 0.68) | 0.70 (0.70 - 0.70) |
| | EfficientNet-B4 | 0.61 (0.60 - 0.61) | 0.68 (0.68 - 0.68) | 0.68 (0.68 - 0.68) | 0.70 (0.70 - 0.70) | 0.71 (0.71 - 0.71) |
| | ResNet-152 | 0.58 (0.58 - 0.58) | 0.69 (0.69 - 0.69) | 0.70 (0.70 - 0.70) | 0.71 (0.71 - 0.71) | 0.72 (0.72 - 0.72) |
| | **HeartBEiT** | **0.64 (0.64 - 0.64)** | **0.73 (0.73 - 0.73)** | **0.74 (0.74 - 0.74)** | **0.74 (0.74 - 0.74)** | **0.75 (0.75 - 0.75)** |
| | Testing samples | 13,859 | 13,859 | 13,859 | 13,859 | 13,859 |
| | **ST-Elevation Myocardial Infarction** | | | | | |
| Internal testing | ViT-B/16 | 0.27 (0.26 - 0.27) | 0.26 (0.25 - 0.26) | 0.20 (0.20 - 0.21) | 0.15 (0.14 - 0.15) | 0.17 (0.17 - 0.17) |
| | EfficientNet-B4 | 0.40 (0.40 - 0.41) | 0.54 (0.53 - 0.54) | 0.60 (0.60 - 0.61) | 0.62 (0.61 - 0.62) | 0.64 (0.64 - 0.65) |
| | ResNet-152 | 0.47 (0.46 - 0.47) | 0.56 (0.55 - 0.56) | 0.58 (0.58 - 0.58) | 0.59 (0.59 - 0.59) | **0.68 (0.67 - 0.68)** |

| | HeartBEiT | 0.56<br>(0.56 - 0.56) | 0.58<br>(0.58 - 0.59) | 0.65<br>(0.65 - 0.65) | 0.64<br>(0.64 - 0.65) | 0.67<br>(0.66 - 0.67) |
|---|---|---|---|---|---|---|
| | Fine-tuning samples | 174 | 1,744 | 4,362 | 8,724 | 17,449 |
| | Testing samples | 4,352 | 4,352 | 4,352 | 4,352 | 4,352 |

**Table 3.** Area Under Precision Recall Curve (AUPRC) metrics for various models compared to HeartBEiT at different fractions of fine-tuning and testing data

**Supplementary Materials**

**Figure 1.** Pretraining curves for HeartBEiT.

**Figure 2.** Receiver Operating Characteristics Curves for Internal Testing of Vit-B/16, EfficientNet-B4, ResNet-152, and HeartBEiT for Classification of Hypertrophic Cardiomyopathy, Left Ventricular Ejection Fraction ≤40%, and ST-Elevation Myocardial Infarction.

**Figure 3.** Precision-Recall Curves for Internal Testing of Vit-B/16, EfficientNet-B4, ResNet-152, and HeartBEiT for classification of Hypertrophic Cardiomyopathy, Left Ventricular Ejection Fraction ≤40%, and ST-Elevation Myocardial Infarction.

**Figure 4.** Receiver Operating Characteristics Curves for External Validation of Vit-B/16, EfficientNet-B4, ResNet-152, and HeartBEiT for Classification of Hypertrophic Cardiomyopathy, Left Ventricular Ejection Fraction ≤40%, and ST-Elevation Myocardial Infarction.

**Figure 5.** Precision-Recall Curves for External Validation of Vit-B/16, EfficientNet-B4, ResNet-152, and HeartBEiT for Classification of Hypertrophic Cardiomyopathy, Left Ventricular Ejection Fraction ≤40%, and ST-Elevation Myocardial Infarction.

**Figure 6 a/b.** Gradient-Weighted Class Activation Saliency Mapping Images for ECG Plots for Classification of Left Ventricular Ejection Fraction ≤40% at 1% and 100% of training data.

**Figure 7 a/b.** Gradient-Weighted Class Activation Saliency Mapping Images for ECG Plots for Classification of Hypertrophic Cardiomyopathy at 1% and 100% of Training Data.

**Figure 8.** Gradient-Weighted Class Activation Saliency Mapping Images for ECG Plots for Classification of ST-Elevation Myocardial Infarction at 100% of Training Data.

**Figure 9.** Average Pairwise Wasserstein Distance Across Data Modalities.

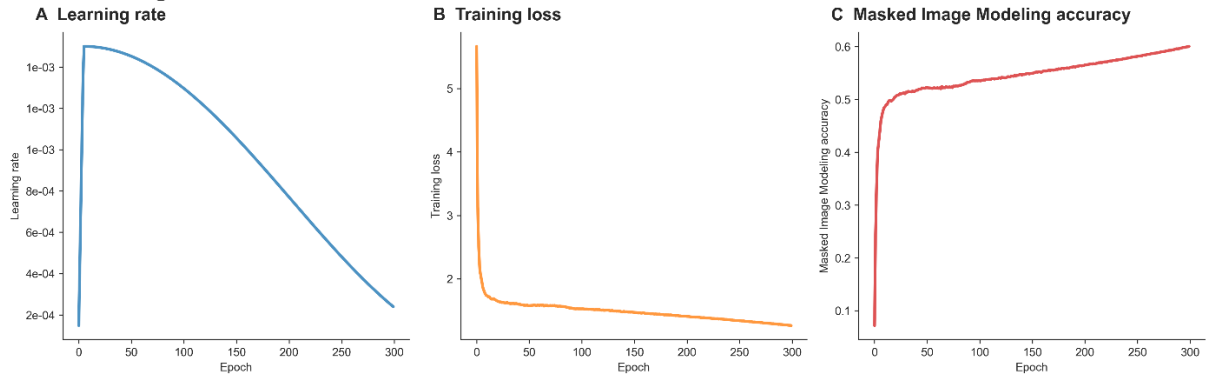

**Figure 1.** Pretraining curves for HeartBEiT.

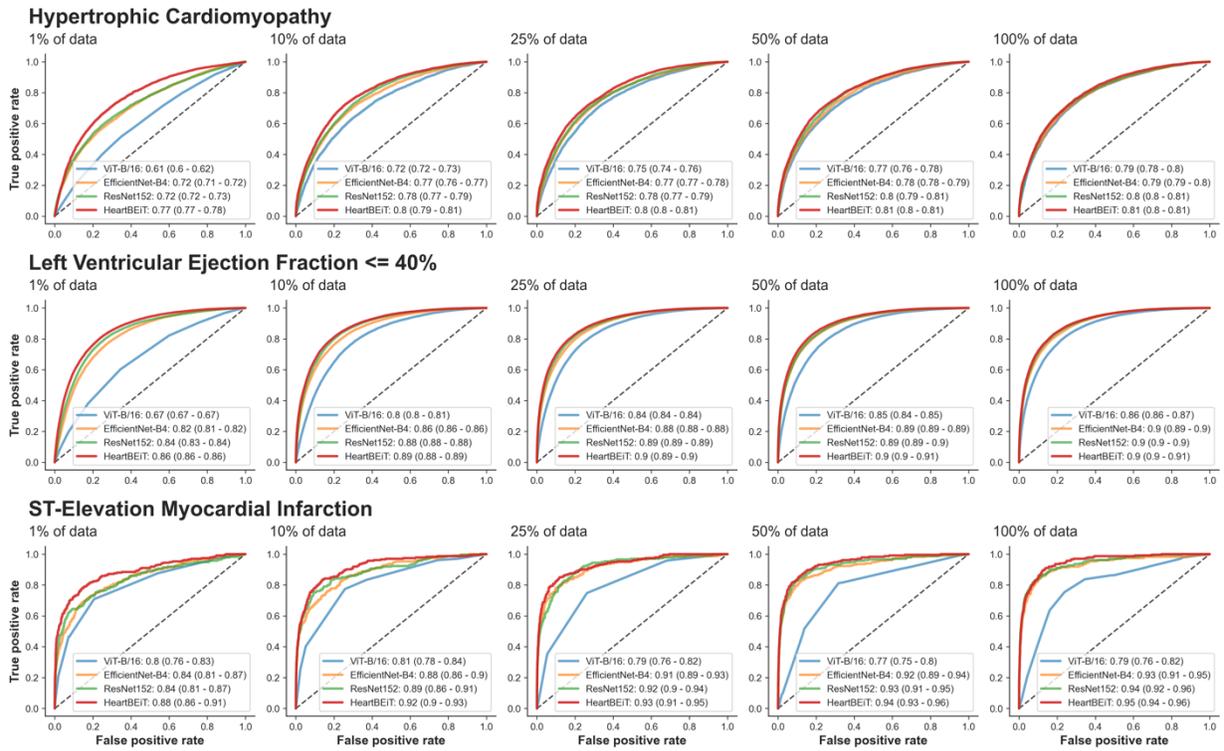

**Figure 2.** Receiver Operating Characteristics Curves for Internal Testing of Vit-B/16, EfficientNet-B4, ResNet-152, and HeartBEiT for classification of Hypertrophic Cardiomyopathy, Left Ventricular Ejection Fraction ≤40%, and ST-Elevation Myocardial Infarction.

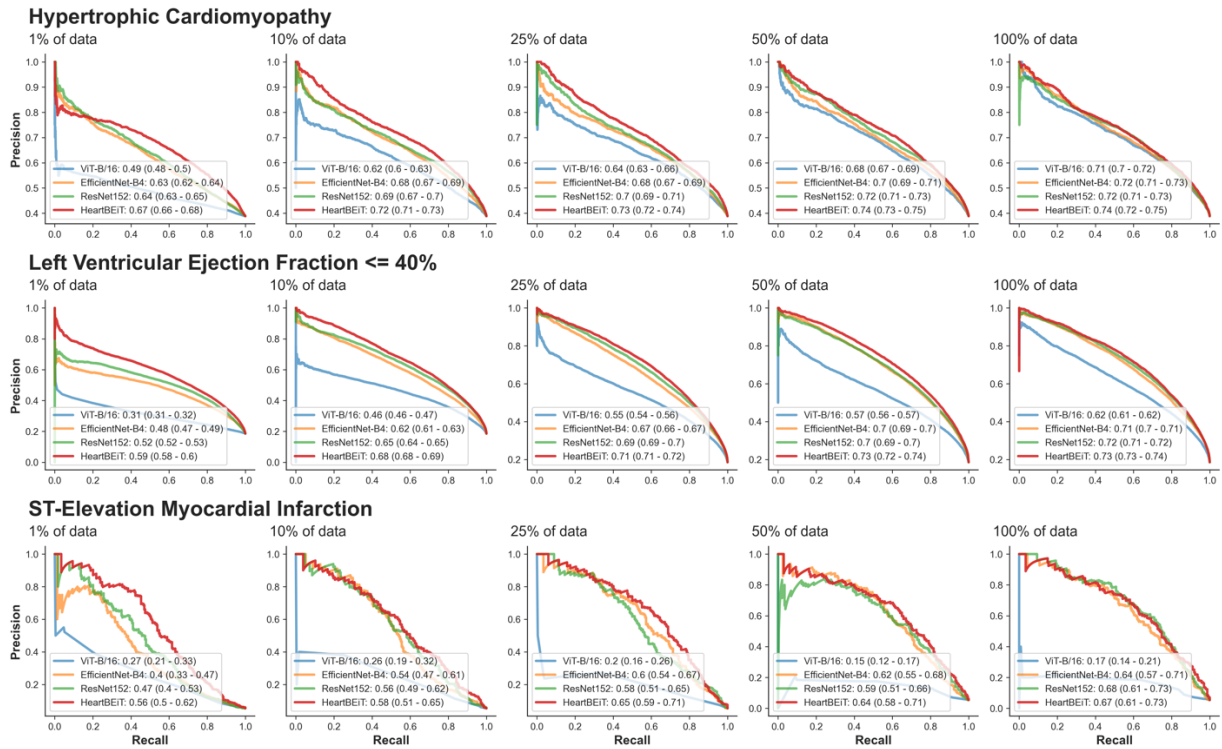

**Figure 3.** Precision-Recall Curves for Internal Testing of Vit-B/16, EfficientNet-B4, ResNet-152, and HeartBEiT for classification of Hypertrophic Cardiomyopathy, Left Ventricular Ejection Fraction ≤40%, and ST-Elevation Myocardial Infarction.

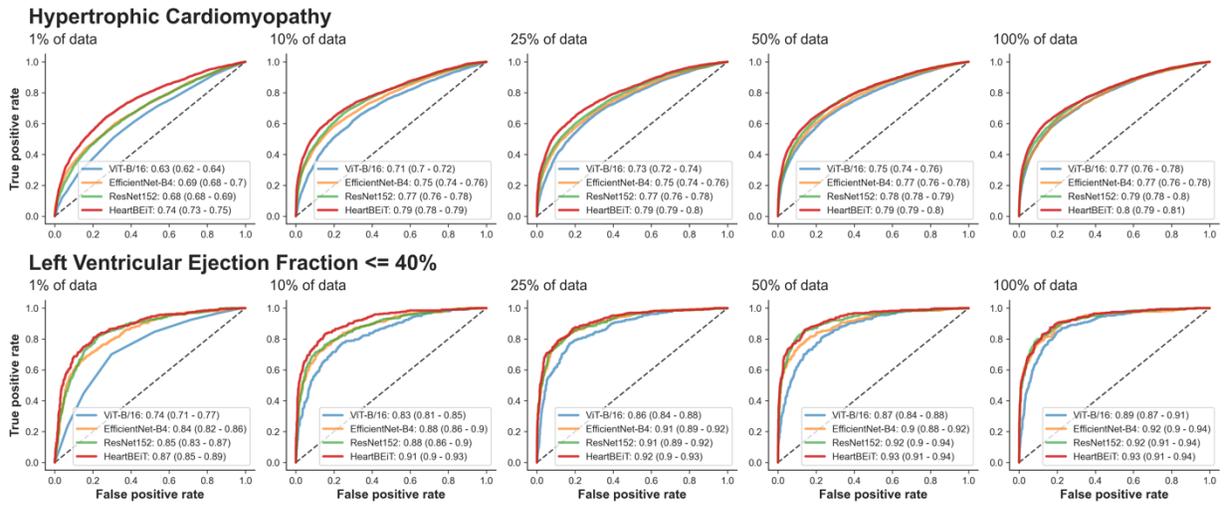

**Figure 4.** Receiver Operating Characteristics Curves for External Validation of Vit-B/16, EfficientNet-B4, ResNet-152, and HeartBEiT for classification of Hypertrophic Cardiomyopathy, Left Ventricular Ejection Fraction ≤40%, and ST-Elevation Myocardial Infarction.

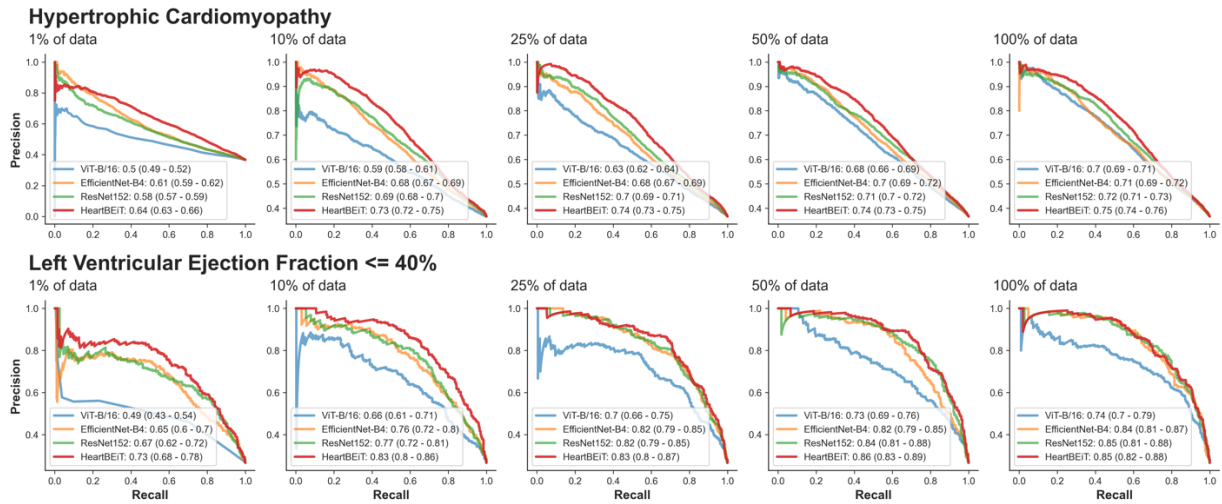

**Figure 5.** Precision-Recall Curves for External Validation of Vit-B/16, EfficientNet-B4, ResNet-152, and HeartBEiT for classification of Hypertrophic Cardiomyopathy, Left Ventricular Ejection Fraction ≤40%, and ST-Elevation Myocardial Infarction.

# Left Ventricular Ejection Fraction <= 40%

Fraction of training data: 0.01

### A ViT-B/16
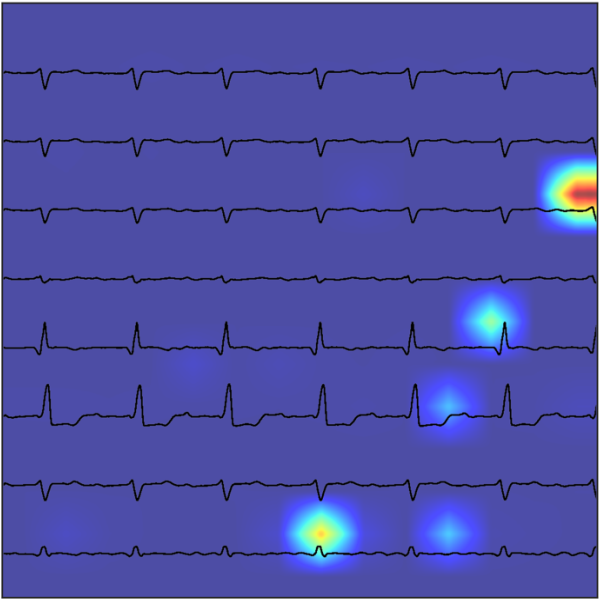

### B EfficientNet-B4
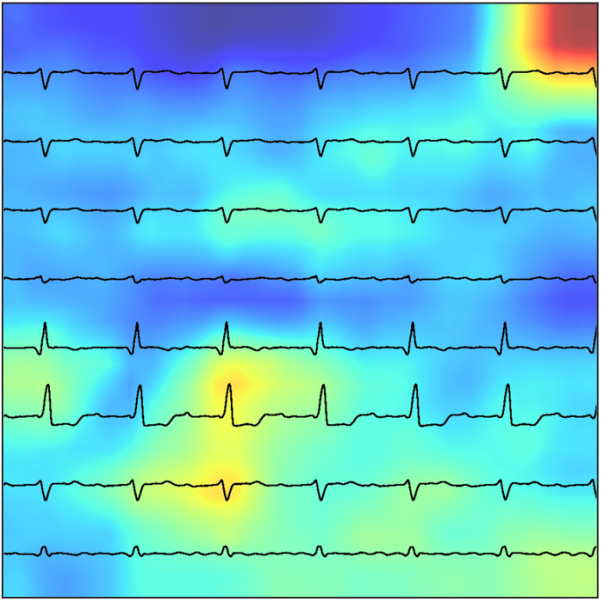

### C ResNet152
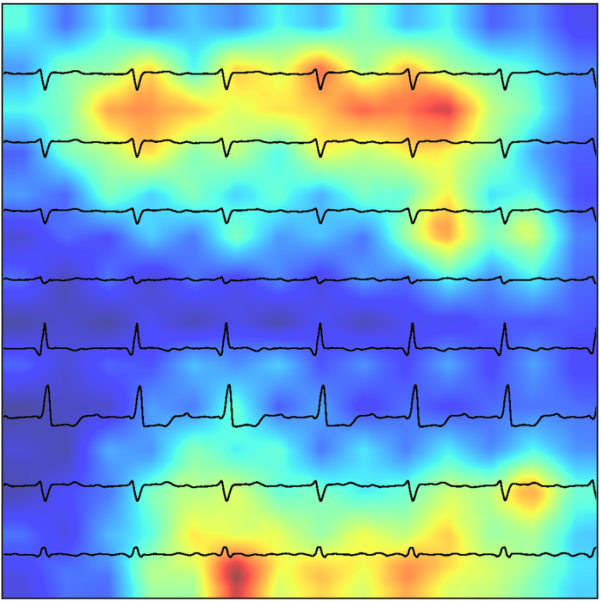

### D HeartBEiT
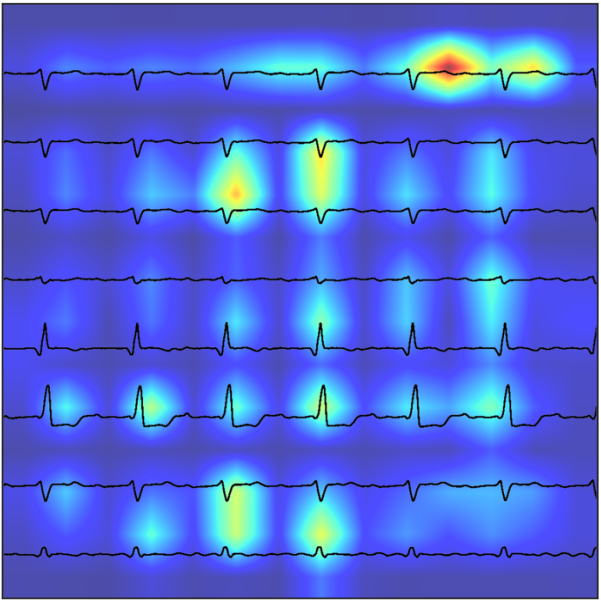

**Figure 6a.** Gradient-weighted class activation saliency mapping images for ECG plots for classification of Left Ventricular Ejection Fraction ≤40% at 1% of training data.

## Left Ventricular Ejection Fraction <= 40%

Fraction of training data: 1.00

**A** ViT-B/16

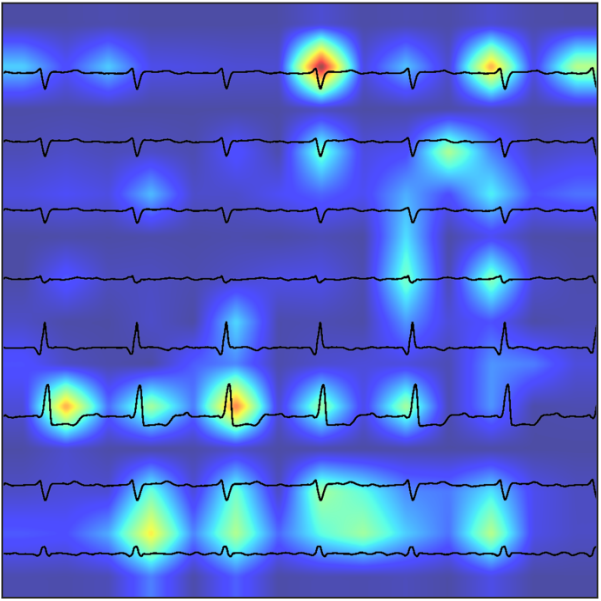

**B** EfficientNet-B4

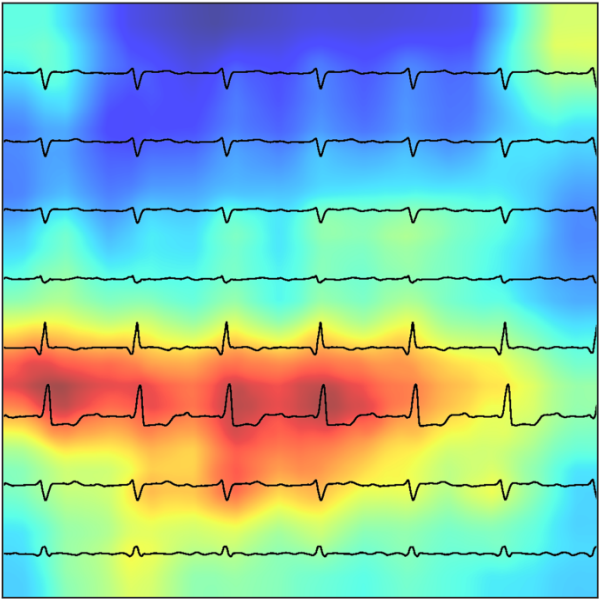

**C** ResNet152

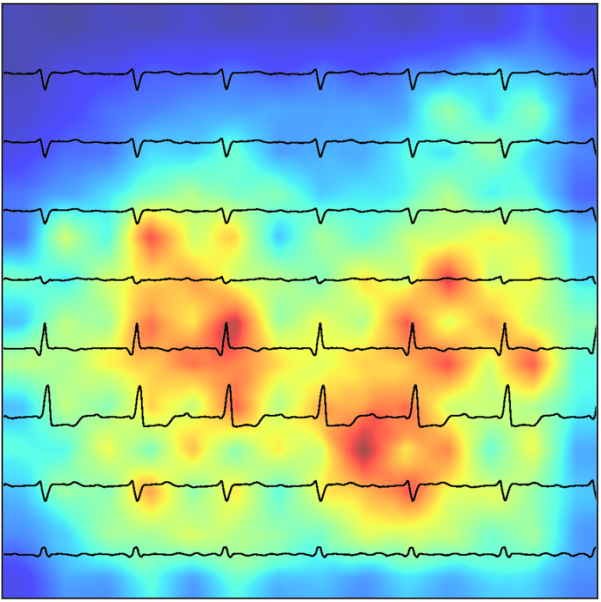

**D** HeartBEiT

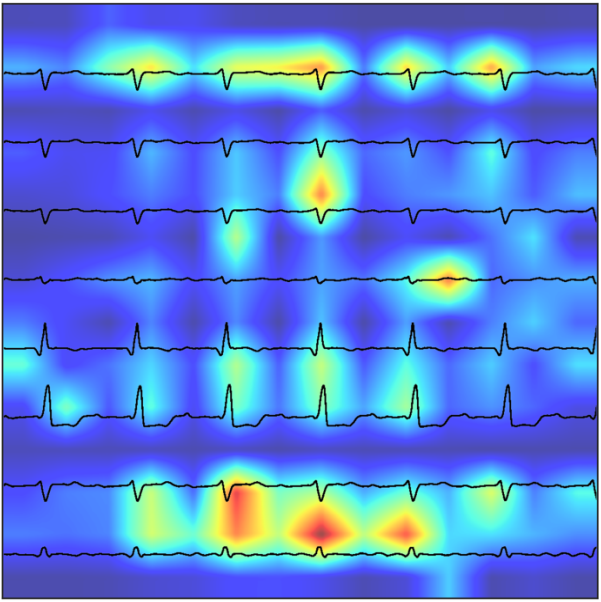

**Figure 6b.** Gradient-weighted class activation saliency mapping images for ECG plots for classification of Left Ventricular Ejection Fraction ≤40% at 100% of training data.

# Hypertrophic Cardiomyopathy

Fraction of training data: 0.01

**A** ViT-B/16

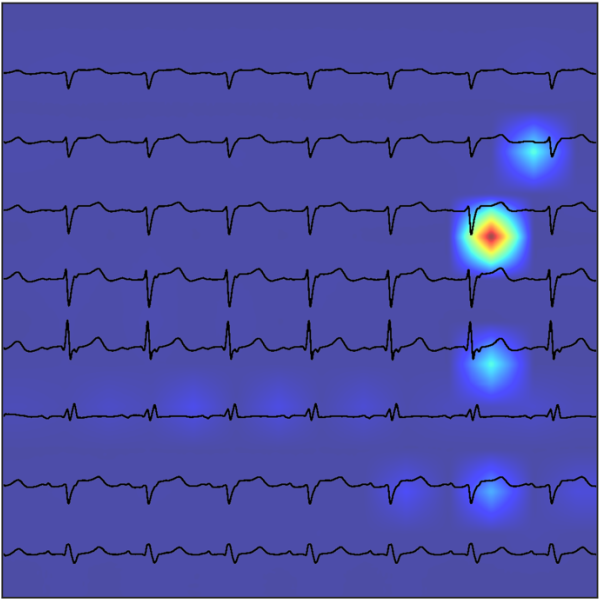

**B** EfficientNet-B4

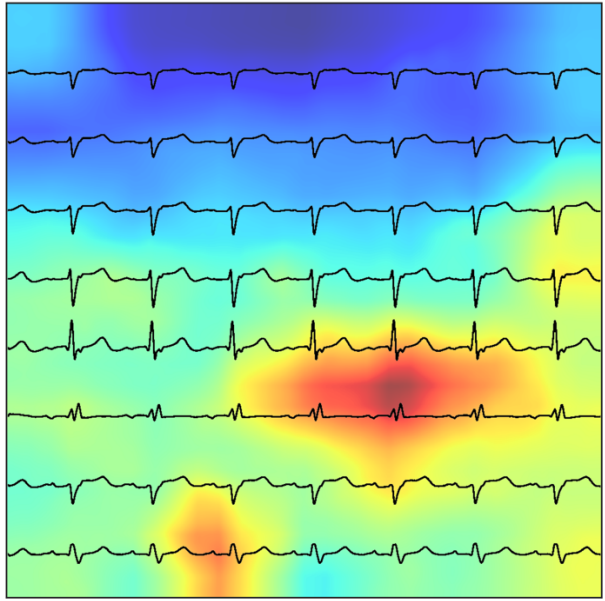

**C** ResNet152

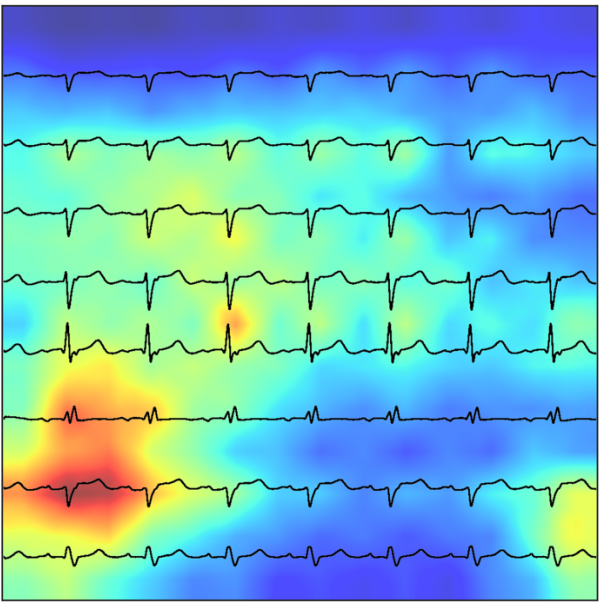

**D** HeartBEiT

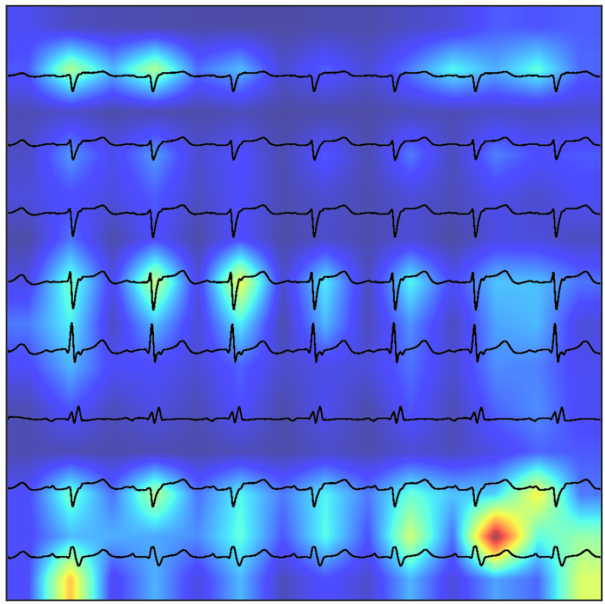

**Figure 7a.** Gradient-weighted class activation saliency mapping images for ECG plots for classification of Hypertrophic Cardiomyopathy at 1% of training data.

# Hypertrophic Cardiomyopathy

Fraction of training data: 1.00

**A** ViT-B/16

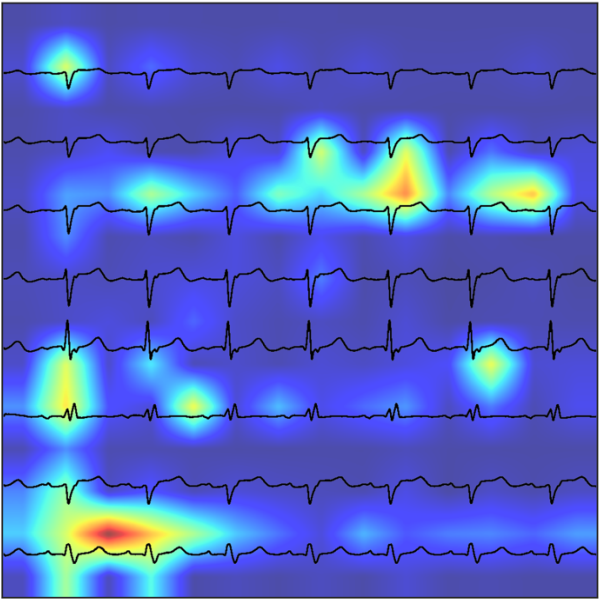

**B** EfficientNet-B4

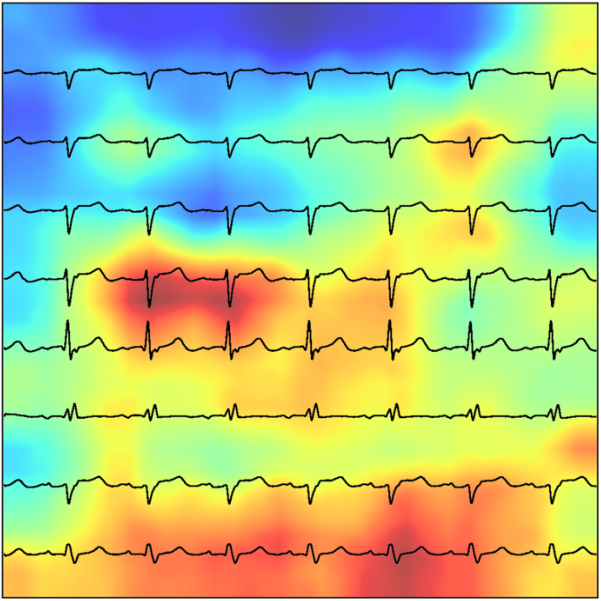

**C** ResNet152

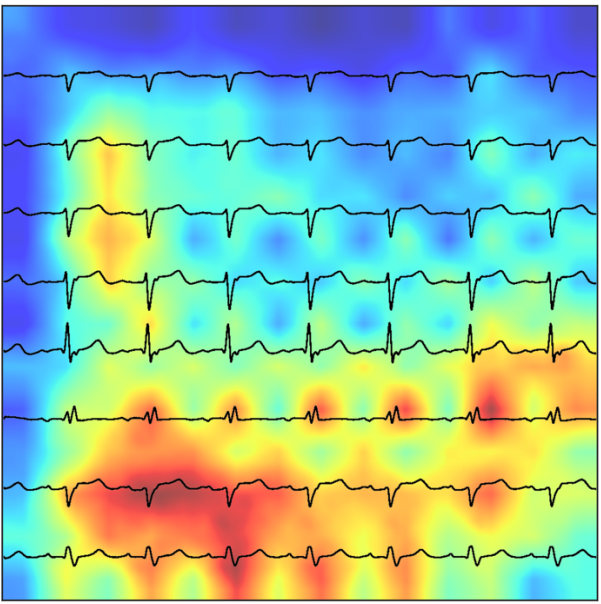

**D** HeartBEiT

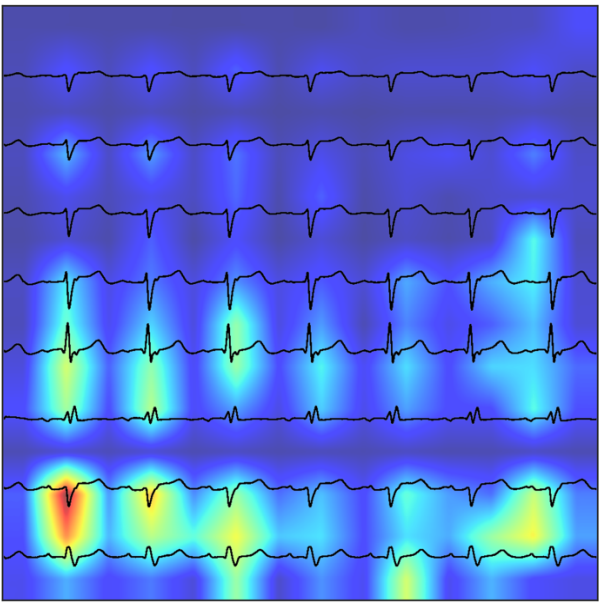

**Figure 7b.** Gradient-weighted class activation saliency mapping images for ECG plots for classification of Hypertrophic Cardiomyopathy at 100% of training data.

## ST-Elevation Myocardial Infarction

Fraction of training data: 1.00

**A** ViT-B/16
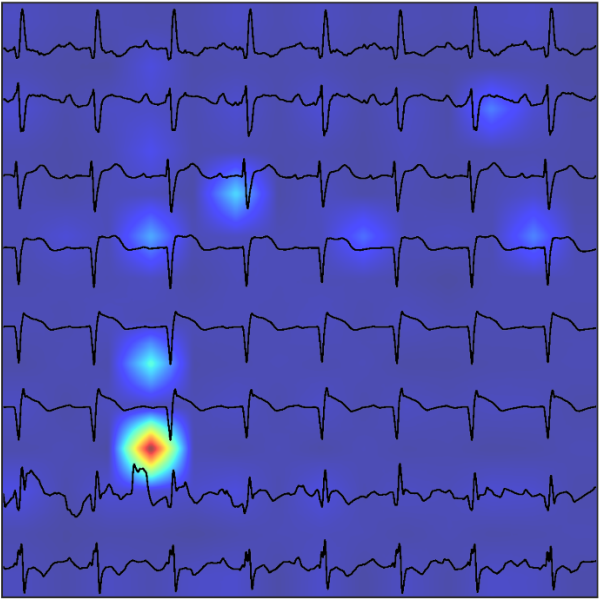

**B** EfficientNet-B4
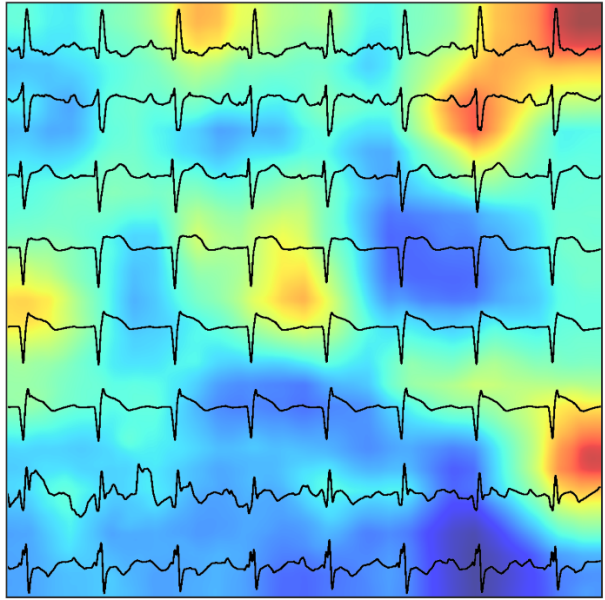

**C** ResNet152
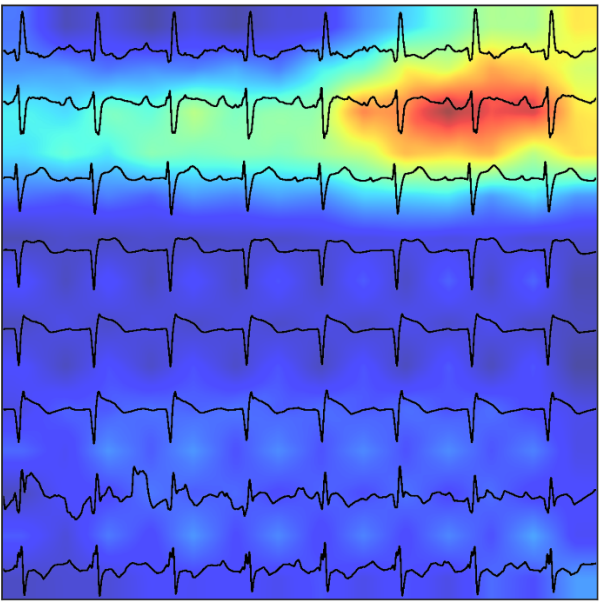

**D** HeartBEiT
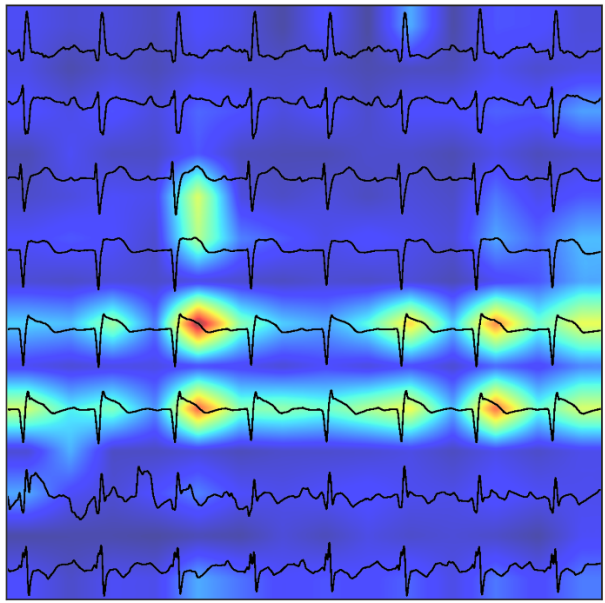

**Figure 8.** Gradient-weighted class activation saliency mapping images for ECG plots for classification of ST-Elevation Myocardial Infarction at 100% of training data.

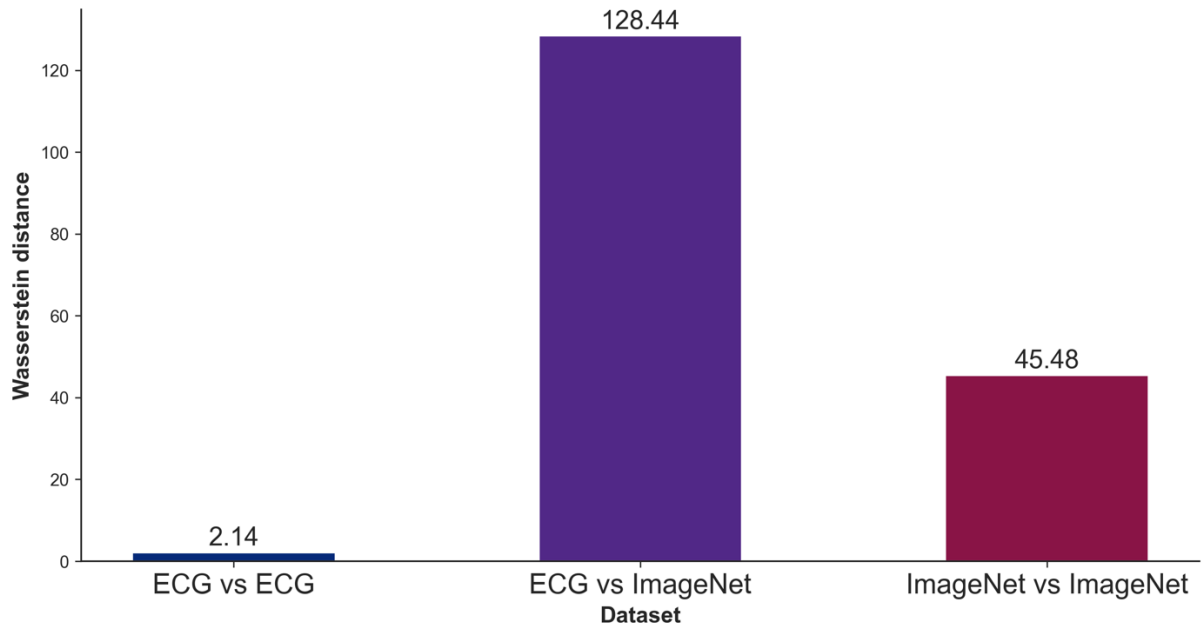

**Figure 9.** Average pairwise Wasserstein distance across data modalities
Metric was calculated by randomly sampling 1000 ECGs and 1000 ImageNet images, and calculating the average pairwise Wasserstein distance between them, such that a total of $10^6$ such calculations were made and averaged.